%% file: 1-main.tex
\documentclass[a4paper,twoside]{article}

\usepackage{calc}
\usepackage{amssymb}
\usepackage{amstext}
\usepackage{amsmath}
\usepackage{amsthm}
\usepackage{multicol}
\usepackage{pslatex}
\usepackage{apalike}
\usepackage{epsfig}

\usepackage[table,xcdraw]{xcolor}
\usepackage{cite}
\usepackage[ruled,norelsize]{algorithm2e}
\makeatletter
\newcommand{\removelatexerror}{\let\@latex@error\@gobble}
\makeatother
\usepackage{array}
\usepackage{algpseudocode}
\usepackage{pifont}
\usepackage{stmaryrd}
\usepackage{mwe}
\usepackage{subcaption}
\usepackage{mdwmath}
\usepackage{mdwtab}
\usepackage{eqparbox}
\usepackage{url}
\usepackage{enumitem}
\usepackage{amsfonts}
\usepackage{makeidx}
\usepackage{epstopdf}
\usepackage{multirow}
\usepackage{alltt}
\usepackage{dsfont}
\usepackage{SCITEPRESS}     

\theoremstyle{definition}
\newtheorem{definition}{Definition}[section]

\newcolumntype{L}[1]{>{\raggedright\let\newline\\\arraybackslash\hspace{0pt}}m{#1}}
\newcolumntype{C}[1]{>{\centering\let\newline\\\arraybackslash\hspace{0pt}}m{#1}}
\newcolumntype{R}[1]{>{\raggedleft\let\newline\\\arraybackslash\hspace{0pt}}m{#1}}

\begin{document}

\title{FPGA Implementation of $\mathbb{F}_2$-Linear Pseudorandom Number Generators Based on Zynq MPSoC: a Chaotic Iterations Post Processing Case Study}

\author{\authorname{Mohammed Bakiri\sup{1,2}, Jean-Fran\c cois Couchot\sup{1}, and Christophe Guyeux\sup{1}}
\affiliation{\sup{1}FEMTO-ST Institute, University of Franche-Comt\'e, Rue du Mar\'echal Juin, Belfort, France}
\affiliation{\sup{2}Centre de D\'eveloppement des Technologies Avanc\'ees, ASM-IPLS team, Algeria}
\email{\{mbakiri, couchot, cguyeux\}@femto-st.fr, mbakiri@cdta.dz}
}


\abstract{Pseudorandom number generation (PRNG) is a key element in hardware security platforms like field-programmable gate array FPGA circuits. In this article, 18 PRNGs belonging in 4 
families (xorshift, LFSR, TGFSR, and LCG) are physically implemented in a FPGA and compared in terms of area, throughput, and statistical tests. Two flows of conception are used for Register Transfer Level (RTL) and High-level Synthesis (HLS).
Additionally, the relations between linear complexity, 
seeds, and arithmetic operations on the one hand, and the resources deployed in FPGA on the other hand, are deeply investigated.
In order to do that, a SoC based on Zynq EPP with ARM Cortex-$A9$ MPSoC is developed to accelerate the implementation and the tests of various PRNGs on FPGA hardware. A case study is finally proposed using \emph{chaotic iterations} as a post processing for FPGA. The latter has improved the statistical profile of a combination of PRNGs that, without it, failed in the so-called TestU01 statistical battery of tests.}

\onecolumn \maketitle \normalsize \vfill

\section{\uppercase{Introduction}}
\label{sec:introduction}
\input{2-Introduction_new.tex}

\section{\uppercase{$\mathbb{F}_2$-Linear generators}}
\label{sec:back}
\input{3-PRNG_new.tex}

\section{\uppercase{Hardware Implementation}}
\label{sec:prng}
\input{4-FPGA_new.tex}

\section{\uppercase{SoC system Based on Zynq Platform for PRNG}}
\label{sec:Zynq}
\input{5-Zynq_new.tex}

\section{\uppercase{Chaotic Iteration Post Processing}}
\label{sec:comp}
\input{6-Chaos.tex}

\section{\uppercase{Conclusion}}
\label{sec:conclusion}
A novel implementation of various PRNGs in FPGA is detailed in this paper, in which two flows of conception (RTL and HLS) demonstrate the performance level of each PRNG in terms of area throughout and statistical tests. Our study has shown that these performances are related to linear complexity, seed size, and arithmetic operations. In order to investigate these parameters, a SoC based on Zynq EPP platform (hardware and firmware) has been developed to accelerate the implementation and tests of various PRNGs on FPGA. On this platform, xorshift$64$ and LFSR$113$ have outperformed the other candidates when considering hardware performance, while PCG$32$ and xorshift$^*$ are the best when studying statistical ones (they succeeded to pass the whole TestU01 batteries). 
Finally, a hardware post processing treatment based on chaotic iterations has been proposed, which has achieved to improve the statistical profile of flawed generators.
We plan to investigate which combinations and parameters of chaotic iterations can be chosen to reach an ideal PRNG (fast, small, and secure).

\section*{\uppercase{Acknowledgements}}
This work is partially funded by the Labex ACTION program (contract ANR-11-LABX-01-01).

\bibliographystyle{apalike}
{\small
\bibliography{References}}
\end{document}

%% file: 2-Introduction_new.tex
Producing randomness is a common need in many applications such as simulation~\cite{gentle2013random}, numerical analysis~\cite{zepernick2013pseudo}, computer programing, cryptography~\cite{luby1996pseudorandomness}. 
Such generators are usually divided in two categories: ``pseudorandom''
(PRNGs), which use algorithms to deterministically produce numbers that look like random (they pass statistical tests with success), and ``true'' random number generators (TRNGs) that use a physical source of entropy to produce randomness.

Deterministic algorithms of pseudorandom generation can be developed by targeting a specific hardware system, like a Field Programmable Gate Array (FPGA), before automatically deploying it on the hardware architecture by using ad hoc frameworks.
Modern FPGAs allow rapid prototyping to explore various hardware solutions and  accelerate \emph{Time to Market}.
The design methodology on FPGA relies on the use of two high levels of implementation, namely the
\emph{Register Transfer Level} (RTL) flow and the \emph{High Level Synthesis} (HLS)~\cite{cong2011high} one.
The HLS flow enables an automatic synthesis to FPGA support in a high programing level.
It also accelerates the IP creation by enabling C, C++, and SystemC specifications to generate the RTL level for FPGAs implementation.
Conversely, traditional RTL flow summarizes the Hardware Description Language (HDL) using verilog/VHDL languages. 
In fact, many recent papers use HLS flow to accelerate some research study in many applications like in cryptography~\cite{homsirikamol2015hardware}.

A way to solve at least partially such security issues is to rigorously and directly implement PRNGs on FPGAs.
To do so, we studied the main functionalities and complexity that distinguish one PRNG for another, which are: 
LFSR (LFSR113, LFSR258, and LUT-SR), LCG (PCG$32$, MWC$256$, CMWC$4096$, and MRG$32k3a$), TGFRS (Mersenne Twister, Well$512$, and TT$800$), xorshift (xorshift$64$, xorshift$128$, xorshift$^*$, and xorshift$+$), and Cellular Automata generators (\textit{cf.}, Section~\ref{sec:back}).
Then, Section~\ref{sec:prng} presents a deep analysis to identify characteristics and main proprieties that contribute to the hardware performance of each PRNG. 
To do so, we use a Zynq device~\cite{rajagopalan2011xilinx} and the two flows (HLS \& RTL) as support to develop a complete \emph{System on Chip} physical support for hardware PRNG, which is detailed in Section~\ref{sec:Zynq}.
Due to well known limitations of these linear generators in cryptographic applications (\emph{e.g.}, linear complexity as described in  Section~\ref{sec:prng}), chaotic iterations 
are finally introduced in Section~\ref{sec:comp} as a possible post processing for hardware PRNGs. The latter improves the statistical profile of the generated numbers as verified by the so-called TestU01 battery of tests~\cite{l2007testu01}. 

%% file: 3-PRNG_new.tex
Let $\mathbb{F}_2$ be the finite field of cardinality 2. Let us firstly recall that a common way to define a pseudorandom number generator is to consider two functions, namely $f:\mathbb{F}_2^\mathsf{N} \rightarrow \mathbb{F}_2^\mathsf{N}$ and $g:\mathbb{F}_2^\mathsf{N} \rightarrow \mathbb{F}_2^\mathsf{M}$, where usually $\mathsf{N} > \mathsf{M}$ and $g$ is one way, such that internally $x_{n+1} = f(x_n)$ is computed, while externally $y_{n+1} = g(x_{n+1})$ is produced ($x_0$ being a seed provided by the user).
A linear PRNG of $r$ bits are a special case of linear recurrence modulo $2$, which can be defined by the following equations: 
\begin{equation}
\centering
\begin{array}{l}
\begin{array}{ll}
x_i = A \times x_{i-1} \ \ (a) & y_i = B \times x_i \ \  (b) 
\end{array}\\
r = {\overset{k}{\underset{\ell= l}{\sum}}} \ y_{\{i,\ell -1\}} \ 2^{-\ell} = y_{\{i,0\}} \  y_{\{i,1\}} \ y_{\{i,2\}} \dots  (c) 
\end{array}
\label{eq:recr}
\end{equation}
Indeed the first equation $(a)$ defines the function $f$, where 
$x_i = (x_{i,0}, \dots , x_{i,{k-1}})$
$\in \mathbb{F}_2^k$ is the $k$-bit vector at step $i$ and $A$ is a $k \times k$ transition matrix with $k$-bit $\mathbb{F}_2$-vector.
The other equations $(b)$ and $(c)$ define the function $g$, where
$y_i = (y_{i,0}, \dots , y_{i,{w-1}})$ 
$\in \mathbb{F}_2^k$ is the $w$-bit output vector at step $i$, while $B$ is a $w \times k$ output transformation matrix with elements in $\mathbb{F}_2$.
The latter produces
the output bits that correspond to the internal RNG state, which is rewritten as
$r \in [0,1]$: the output at step $i$. 
We focus on implementing four families of generators in one or both flows, which are:

\textbf{Linear Feedback Shift Register}. It 
uses a sequence of shift registers to generate one bit per iteration. In such a PRNG, the matrix $A$ represents the LFSR coefficients. Accordingly, if any of these coefficients exists, it deploys a \textit{XOR} operand on some designed registers to build a feeadback input to the first register.
LFSR$113$, LFSR$258$~\cite{l1999tables}, and Taus$88$~\cite{l1996maximally} 
are examples of LFSR.
Additionally, \emph{Look-up Table Shift Register} (LUT-SR)~\cite{thomas2013lut}) is another LFSR generator,
which uses LUT as a $k$-bit shift-register to allow the cascading for any required size. 

\textbf{Linear Congruential Generators}. They 
are based on linear recurrence equations having the form: $x_{i+1} = (ax_{i} + c) \mod 2^k$.
\emph{Multiply-With-Carry} MWC$256$ and \emph{Complementary MWC} CMWC$4096$~\cite{couture1997distribution} 
are two implementations of LCG, where in MWC the increment $c= \lfloor (ax_{i-r} + c_{i-1})/2^k \rfloor$ is an initial carry, 
and the CMWC takes the complement of $(2^k-1)-x_i(MWC)$ to form a new output.
Another example is a new improvement of LCG named PCG$32$~\cite{citation-0}, which uses a permutation function (dropping bits using fixed and random rotations). 
We can also evoke the MRG$32K3a$ generator~\cite{l1999good}, which is a combined \emph{Multiple Recursive Generator} computed as follows: $y_i=x_i/2^k$.

\textbf{Twisted Generalized Feedback Shift Register}. It is based on matrix linear recurrence of $n$ sequence words, each containing $w$-bits. For each recurrence operation $k$,
$k=0,1, \dots,m$, 
the TGFSR operates with three sequence words: the first two sequence words $x_k$ and $x_{k+1}$ being computed with bitmask vectors ($S_{MSB},S_{LSB}$) with the middle sequence word $x_{k+m}$, $0\leqslant m\leqslant n$, as follows: 
\begin{equation}
x_{k+n} = x_{k+m} \oplus  (((x_k \& \underline{S_{MSB}})\ |\ (x_{k+1} \& \overline{S_{LSB}})) \times A).
 \label{eq:Mersenne}
\end{equation}

At iteration $i=k+n$, TGFSR uses a tampering module (bitwise/shift computation) to reduce the dimensionality $n$ of equidistribution.
Mersenne Twister (MT)~\cite{matsumoto1998mersenne}, Well$512$~\cite{panneton2006improved}, and TT800~\cite{matsumoto1994twisted} are examples of TGFSR.

\textbf{XORshift Generators}. They 
are very fast PRNGs, in which the internal state is repeatedly changed by applying a series of shift and exclusive-or ($\textit{XOR}$ $\otimes$) operations.
XORshift$^*$ generators~\cite{vigna2014experimental}, XORshift$64$~\cite{marsaglia2003XORshift}, and XORshift$+$~\cite{vigna2014further} are instances of such generators.

\textbf{Cellular Automata Generator}. This is a discrete generator proposed as formal models of self-reproducing robots. It includes at least $3$ cells with an internal state machine that can be a Boolean function rule. 
Therefore, the CA structure can hold and update the internal state for each cell, depending on the local rules registered by the \emph{Wolfram} code~\cite{gleick1997chaos} ($2^8$ possibilities) and the states of their neighborhoods. 

%% file: 4-FPGA_new.tex
%
%
%
\begin{figure}[!ht]
\centering
\begin{subfigure}[b]{0.43\textwidth}
    \includegraphics[width=\textwidth]{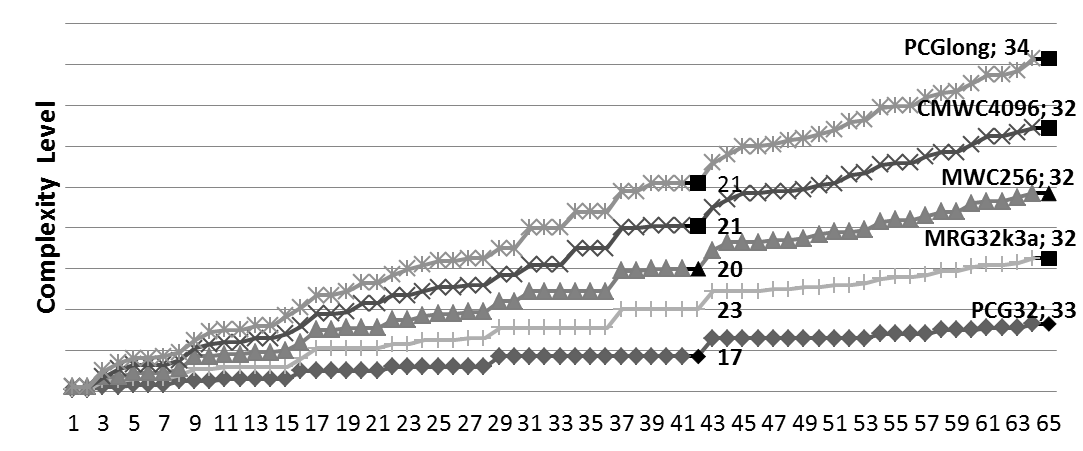}
    \subcaption[short for lof]{LCG Familly}
    \label{fig:subfig1}
\end{subfigure} \vfill
\begin{subfigure}[b]{0.43\textwidth}
    \includegraphics[width=\textwidth]{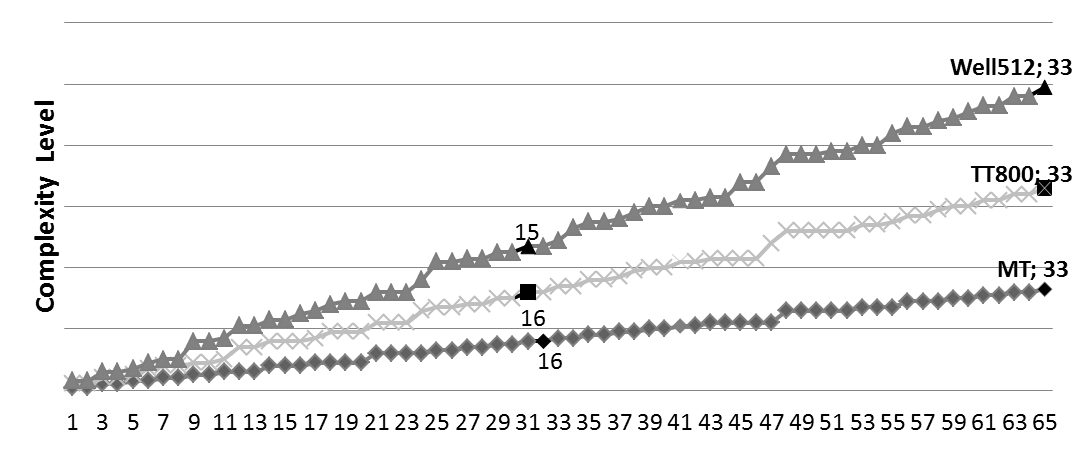}
    \subcaption[short for lof]{TGFSR Familly}
    \label{fig:subfig2}
\end{subfigure}\vfill
\begin{subfigure}[b]{0.43\textwidth}
    \includegraphics[width=\textwidth]{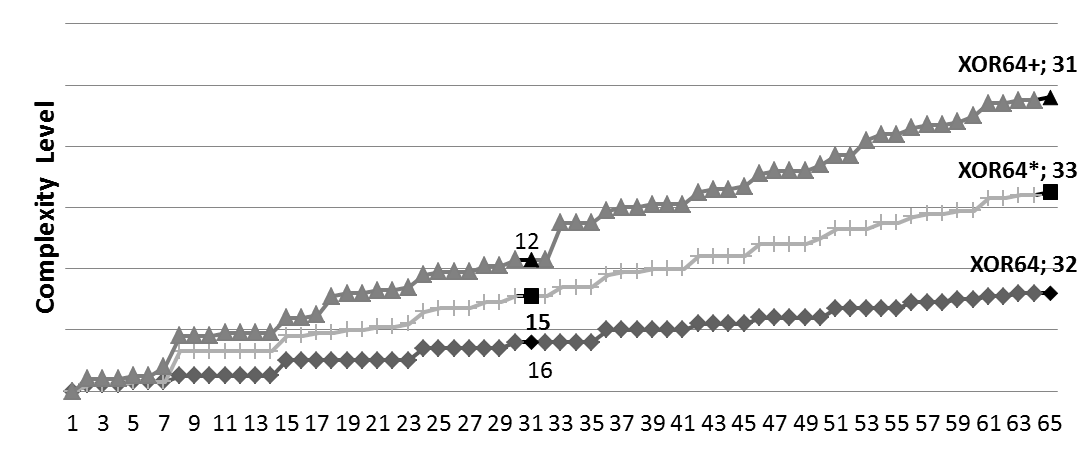}
    \subcaption[short for lof]{xorshift Familly}
    \label{fig:subfig3}
\end{subfigure}\vfill
\begin{subfigure}[b]{0.43\textwidth}
    \includegraphics[width=\textwidth]{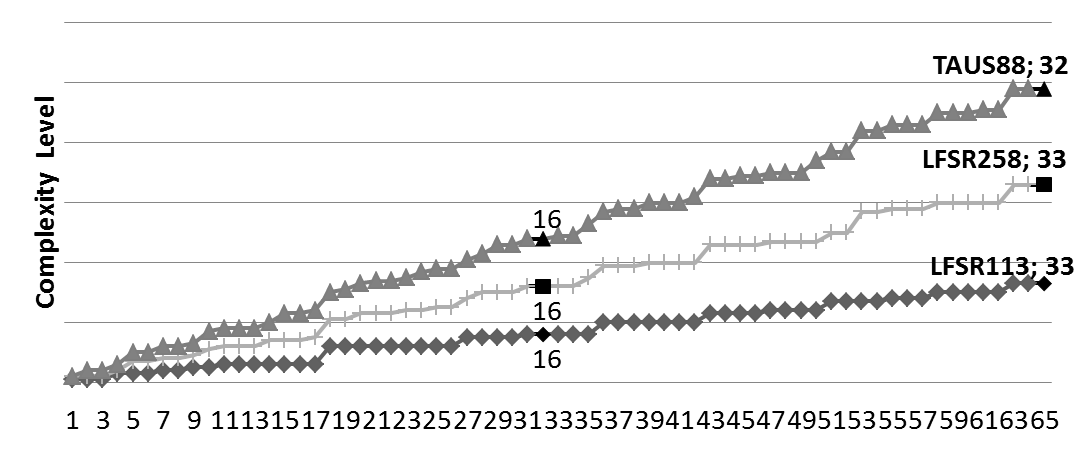}
    \subcaption[short for lof]{LFSR Familly}
    \label{fig:subfig4}
\end{subfigure}
\caption{Computational Complexity Analysis with Berlekamp-Massey Algorithm}
\label{fig:prngcomp}
\end{figure}

In this section, we start a deep analysis of the PRNG implementations on FPGA using \emph{Register Transfer Level} (RTL) and/or \emph{High Level Synthesis} (HLS) flows. Results are studied according to: (1) the space, timing, and computational complexity, (2) the seed and period, and (3) the arithmetic operators and dynamic range FPGA resources. 
Table~\ref{Tab:hlsimp} and Table~\ref{Tab:rtlimp} show obtained results  when implementing $18$ PRNGs. Figure~\ref{fig:prngcomp} presents, for its part, the computation complexity and its impact on performance. Each PRNG is implemented either in just one or both (HLS \& RTL) flows. 
Concerning the software platform, we used Vivado HLS tool for HLS flow and Vivado synthesis for RTL flow of Xilinx.

\begin{table*}[ht]
\centering
\caption{HLS Implementation}
\label{Tab:hlsimp}
\resizebox{\textwidth}{12mm}{%
\begin{tabular}{|l||l|l|l|l|l|l|l|l|l|l|l|l|l|l|}
\hline
PRNG & LFSR113 & TAUS88 & PCG32 & MRG32k3a & TT800 & WELL512 & MWC256 & CMWC4096 & XOR$^*$ & LFSR258 & XORP128 & XORP64 & XOR$+$ & KISS124 \\ \hline \hline
Output Range & 32 & 32 & 32 & 32 & 32 & 32 & 32 & 32 & 64 & 64 & 64 & 64 & 64 & 64 \\ \hline
Period 2\textasciicircum  & 113 & 88 & 32 & 191 & 800 & 512 & 8222 & 131086 & 1024 & 258 & 128 & 64 & 128 & 124 \\ \hline
LUT & 66 & 56 & 371 & 214 & 173 & 90 & 219 & 285 & 303 & 132 & 49 & 64 & 136 & 271 \\ \hline
FF & 113 & 88 & 367 & 522 & 549 & 147 & 399 & 471 & 394 & 258 & 64 & 65 & 133 & 746 \\ \hline
RAM & 0 & 0 & 0 & 0 & 2 & 2 & 1 & 8 & 4 & 0 & 0 & 0 & 4 & 0 \\ \hline
DSP & 0 & 0 & 10 & 8 & 6 & 0 & 4 & 2 & 10 & 0 & 0 & 0 & 0 & 7 \\ \hline
Frequences Mhz & 769 & 555 & 333 & 160 & 160 & 214 & 153 & 148 & 224 & 617 & 510 & 894.45 & 225 & 149 \\ \hline
Area & 1432 & 1152 & 5904 & 5888 & 5776 & 1896 & 4944 & 6048 & 5576 & 3120 & 904 & 1032 & 2152 & 8136 \\ \hline
Throughput Gbps & 24.6 & 17.76 & 10.6 & 5.12 & 5.12 & 6.8 & 4.9 & 4.7 & 14.33 & 39.5 & 156.32 & 57.24 & 14.40 & 4.7 \\ \hline
\end{tabular}}
\end{table*}

\begin{table*}[ht]
\centering
\caption{RTL Implementation on FPGA}
\label{Tab:rtlimp}
\resizebox{\textwidth}{12mm}{%
\begin{tabular}{|l||c|c|c|c|c|c|c|c|c|c|c|}
\hline
PRNG & MT\_WS & MT\_NS & LUT-SR & CA & LFSR113 & TAUS88 & LFSR258 & XORP128 & XORP64 & XOR$+$ & KISS124 \\ \hline \hline
Output Rang & 32 & 32 & 32 & 32 & 32 & 32 & 64 & 64 & 64 & 64 & 64 \\ \hline
Period 2\textasciicircum  & 19937 & 19937 & 1024 & 32 & 113 & 88 & 258 & 128 & 64 & 128 & 124 \\ \hline
LUT & 523 & 184 & 64 & 98 & 95 & 96 & 207 & 53 & 65 & 147 & 742 \\ \hline
FF & 120 & 179 & 64 & 40 & 128 & 77 & 320 & 128 & 64 & 196 & 256 \\ \hline
RAM & 2 & 2 & 0 & 0 & 0 & 0 & 0 & 0 & 0 & 0 & 0 \\ \hline
DSP & 3 & 0 & 0 & 0 & 0 & 0 & 0 & 0 & 0 & 0 & 6 \\ \hline
Frequences Mhz & 118 & 462 & 609 & 598 & 595 & 667 & 556 & 531 & 588 & 403 & 78.1 \\ \hline
Area & 5144 & 3272 & 576 & 1104 & 1784 & 1384 & 4216 & 1448 & 1032 & 2744 & 7984 \\ \hline
Throughput Gbps & 3.8 & 13.2 & 19.5 & 19.1 & 19 & 21.3 & 35.5 & 17 & 37.6 & 25.7 & 5 \\ \hline
\end{tabular}}
\end{table*}

\subsection{Space, Timing, and Computational Complexities}
\label{cpmlex}
The space represents the allocated cost of most objects used in the algorithm (tables, indexes, loops, etc.). Regarding FPGAs, the latter can be translated in memories, registers, and LUT resources, etc. 
The question raised in this section is thus: how much space states are needed to provide pseudorandom numbers with a good statistics profile? We wonder too whether there is any relation between the space (mean resources) used in FPGA and a success in passing stringent statistical \emph{Linear Complexity Test}~\cite{blackburn1994aspects} of test. 
To answer this question, we first define what is a linear complexity.

Most PRNGs mentioned in this article are linearly recursive. If we take a finite binary sequence $(x_i) = (x_{i,0}, \dots , x_{i,{k-1}})$ $\in \mathbb{F}_2^k$, its linear complexity $L_{k}(x_i)$ is the length of the shortest characteristic polynomial (see Equation~\eqref{eq:recr}) of the LFSR generating the same sequence (for a sequence equal to $x_0=x_1=\dots=x_{k-2}=0$ and $x_{k-1}=1$, the linear complexity is $k$ and $L_{k+1}>L_k$). Non randomness is claimed when the length is short. 
This is confirmed by the fact that almost all generators (with the exception of PCG$32$, xorshift$^*$, and MRG$32k3a$) presented in this article fail in statistical \emph{Linear Complexity Test} of Test.

A first way to compute this complexity is to consider the NIST tests battery~\cite{Nist10}. But the improved Test battery additionally incorporates some ``jump'' aspects in this test, leading to the fact that most generators succeeding in NIST linear complexity test finally fail to pass the one of Test. Indeed, the latter calculates the jumps that occur in the linear complexity for each local subsequence, that is, the $k$'s that satisfy $L(k)-L(k-1)>0$. This number of jumps represents how much bits have to be added to the sequence to increase its linear complexity. 
Ideal PRNGs have to perform jumps symmetric to the $k/2$-line~\cite{rueppel1985linear}, as in a perfect linear complexity, maximum jump heights of $k/4$ and  close to $\lfloor(k+1)/2\rfloor$ for $k$-sequences are required. 

\begin{figure}[!h]
  \centering
   \includegraphics[width=0.43\textwidth]{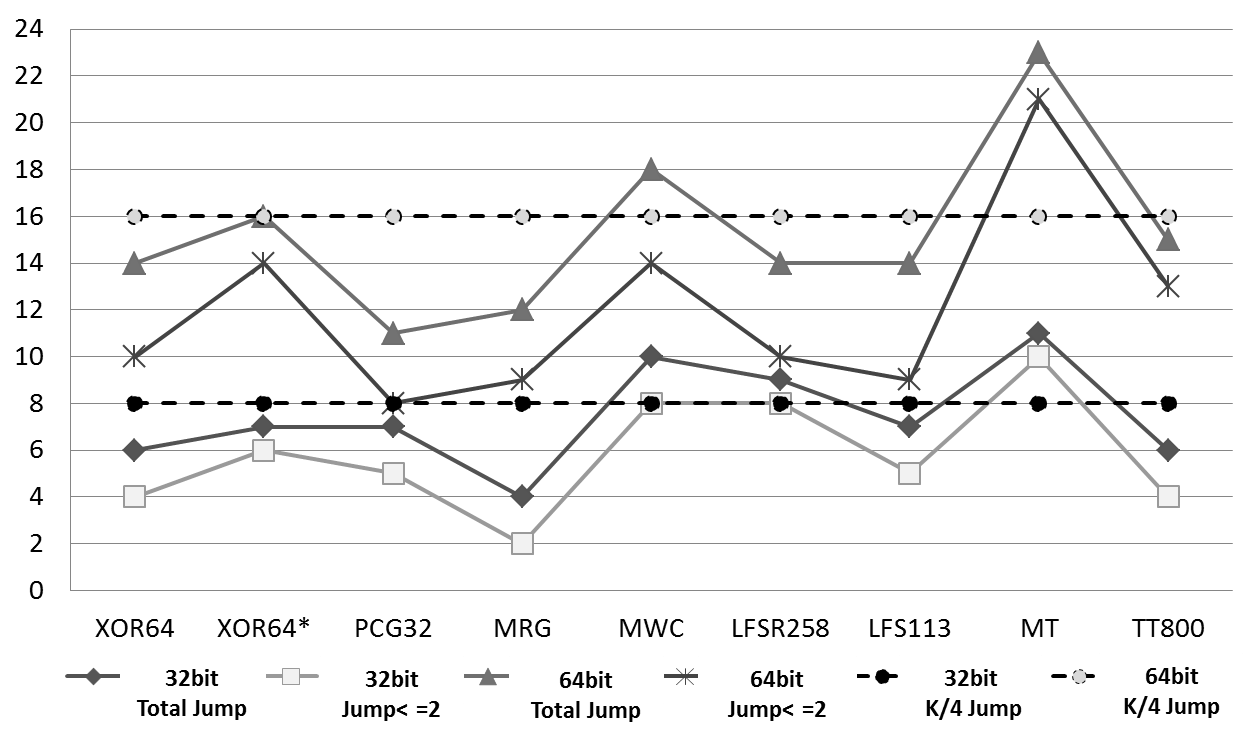}
  \caption{Jump Computation for 32/64 bit of random}
  \label{fig:jumpcomp}
 \end{figure}

Regarding FPGAs, these jumps determine how much resources are required in order to have a perfect complexity profile. 
For illustration purposes, some of these PRNG jumps have been computed, see Figure~\ref{fig:jumpcomp}. 
Concerning $32$ bit sequences, the number of perfect successive jumps ($<2$) is large for all PRNGs (XOR64, for instance, has a total of $6$ jumps, $4$ of them being perfect).
However, in the $64$ bit case, two kind of results have been obtained. 
On the one hand, we found PCG$32$ and MRG that can pass Test have low successive jumps compared to xorshift$^*$. This is due to the multiplication space used for these generators. This is confirmed in Figure~\ref{fig:prngcomp}, that summarizes the linear complexity for each family of PRNGs, which is close to $k/2=32$. 

Let us now consider xorshift$^*$ generators, which also use 64-bit multiplications. 
Their linear complexity is closely perfect, as can be seen in Figure~\ref{fig:prngcomp}. 
The key difference here is the permutation function used for multiplication. 
In LCG family, this is the main function applied to perform an uniform scrambling operation.
On the opposite, they are deployed to inject bias in randomness in xorshift$^*$.
The PCG$32$ deploys 64-bit multiplications (128-bit state), but it uses only 36-bit of state while always dropping the MSB parts (the states space used are constant for any operation).
This fact means a loss of information that can create a new jump in complexity, even if we use more complected seeds (\emph{i.e.}, pcglong). In other words, it needs some time to be perfectly linear (see Figure~\ref{fig:prngcomp}(\subref{fig:subfig1}) starting from 41-bit). In hardware level, doing the same operation leads to unnecessary area and power consuming.

The second point to investigate is the size and number of jumps in complexity profile. 
If we consider multiplications for instance, each PRNGs embedding them needs $2*n$ outputs of multipliers (DSP or LUT blocs in FPGA) for each $n$-bit input multiplication: for each jump, 
an additional input multiplier is used. In other words and compared to stable complexity, a fixed jump during time does not use the full capacity of the multiplier (see Section~\ref{sub:arith}).

\subsection{Seed and Period}
Most generator implementations require a seed to initiate the internal states. It is also a space deterministic parameter for the PRNG. Regardless of the space size, the consumption can be quite large if the seed is large. 
This seed can be: single or multiple value(s) in table(s), a constant or a value generated from a given algorithm, or it can even be extracted from a physical source (TRNG). Additionally, the seed can also contribute to the period of the PRNG. 
A period of a power of two is recommended to have an uniform output, due to the following reason: if it is not the case, some hardware resources cannot be used (\emph{e.g.}, MRG$32$k$3$ 
has an output of $2^{32}-209$ and $209$ values are never used).

In our implementations (RTL and HLS), we choose to seed TGFSR and MWC generators with an array using one of Knuth's generators (see~\cite[p. 106]{Knuth:1997:ACP:270146}  for multiplier). Depending on the seed period and using MT as an example, we can store each value of the seed in one memory at a time and for each clock cycle. The RAM memory, configured in the read-before-write mode, operates like a feedback shift register. 
In this mode, new inputs are stored in memory at an appropriate write address, while the previous data are transferred to the output ports. The latter, coming from RAM, are then processed following the Equation~\eqref{eq:Mersenne}.
Therefore, different address controllers are used for each process (seed and generation). For the other PRNGs,
the seed can be a constant or generated by another algorithm.

Let us illustrate the performance impact using Mersenne Twister (MT) with (WS) and without (NS) the seed algorithm in RTL level. 
When including the seed in implementation, 
we need to store $624$ values in two memories for each clock cycle, which are used later in 
random transformation and tempering. Therefore, the total area and time resources is increased.
Otherwise, in the case of the absence of the seed, the latter is generated and stored  separately in memories, before the deployment of the PRNG. 
During our comparisons of the two approaches on MT generator, we have remarked that, with seed, frequency is reduced to less than $200$MHz compared to the case without it. Therefore, to increase performances, most PRNGs do not include the seed internally (software is used).
The LUT-SR PRNG is an exception, which consumes less space but needs to wait $1,024$ clock cycles for the seed generation.

\subsection{Arithmetic Operators and Dynamic Range}\label{sub:arith}
The arithmetic operators area is a key issue at hardware level, which can be considered as a major factor of the quality of the
final implementation. These operators can be a single basic operation (like addition or subtraction, multiplication of variables or constants), algebraic functions (division, modulo, etc.), or any other elementary function. However, in hardware level, these arithmetic operations (specially the multiplication) are hard coded by the tools (Xilinx) using optimized algorithms for that (\emph{Canonical Signed Digit} (CSD), \emph{Booth recoding}, etc.). 

In the binary field $\mathbb{F}_2$, 
most PRNGs use only positive integer values and fixed point representations in hardware level, while if we take for instance the computing of the partial products, the latter can use only \emph{glue logic} (\emph{i.e.}, AND gates or a series of additions).
These partial products are defined as \emph{Distributed Arithmetic} (DA~\cite{meyer2007digital}), they perform a multiply-and-add operation at the same time using most basic logic elements (LUTs). Their size and performance depend on both the \emph{word length} (addressing the LUT increases the table exponentially) and their binary representations, regarding \emph{dynamic range} and precision.
This word length represents the ratio between the largest and the smallest nonzero and positive number that can be represented (integer), which is expressed as follow: $\textit{DR}_{\textit{fxpt}}=r^n-1$ where $r$ is in binary format (Radix-$2$) and $n$ is the number of digits in fixed-point precision.
 
Modern FPGAs use \emph{Digital Signal Processing} (DSP$48$E$1$) slices to obtain the optimal implementation of these operators and avoid overflows and underflows for complex operations. 
It supports many independent functions including multiply, MAC, magnitude comparator, bit-wise logic functions, etc.
Because multiplications are widely used in PRNGs, they can be implemented with DSP used as a $25$x$18$-bit multiplier, and which can be pipe-lined. In Figure~\ref{fig:prngcomp}, we can see the obvious impact of \textit{DR} on computation complexity, which means that larger \textit{DR} are translated to logic space, operator, and timing. Let us take for instance the LFSR$258$ of \textit{DR}$=2^{64}$, which applies exact logic operators as shift, logic AND, and xorshift. Its complexity is linear with the ``DA'' used when $1<\textit{DR}<16$ bits, otherwise it jumps higher with the use of more complicated logic to operate multiplications (DSP) and store values.


%% file: 5-Zynq_new.tex
\subsection{Hardware and Firmware Design}
Xilinx Zynq-$7000$ \emph{Extensible Processing Platform} (EPP)~\cite{rajagopalan2011xilinx} is a silicon system on chip (SoC) for FPGAs, which has been proposed by Xilinx.  
The latter is defined as \emph{Peripheral System} (PS), which is a sub-system with ARM. The full FPGA, for its part, is the \emph{Programmable Logic} (PL) that is connected with PS through an AXI bus interface. Therefore, and for pseudorandom number generation, we have developed a complete SoC infrastructure divided in two parts: hardware and firmware.

\begin{figure}[!ht]
  \centering
  \includegraphics[width=0.46\textwidth]{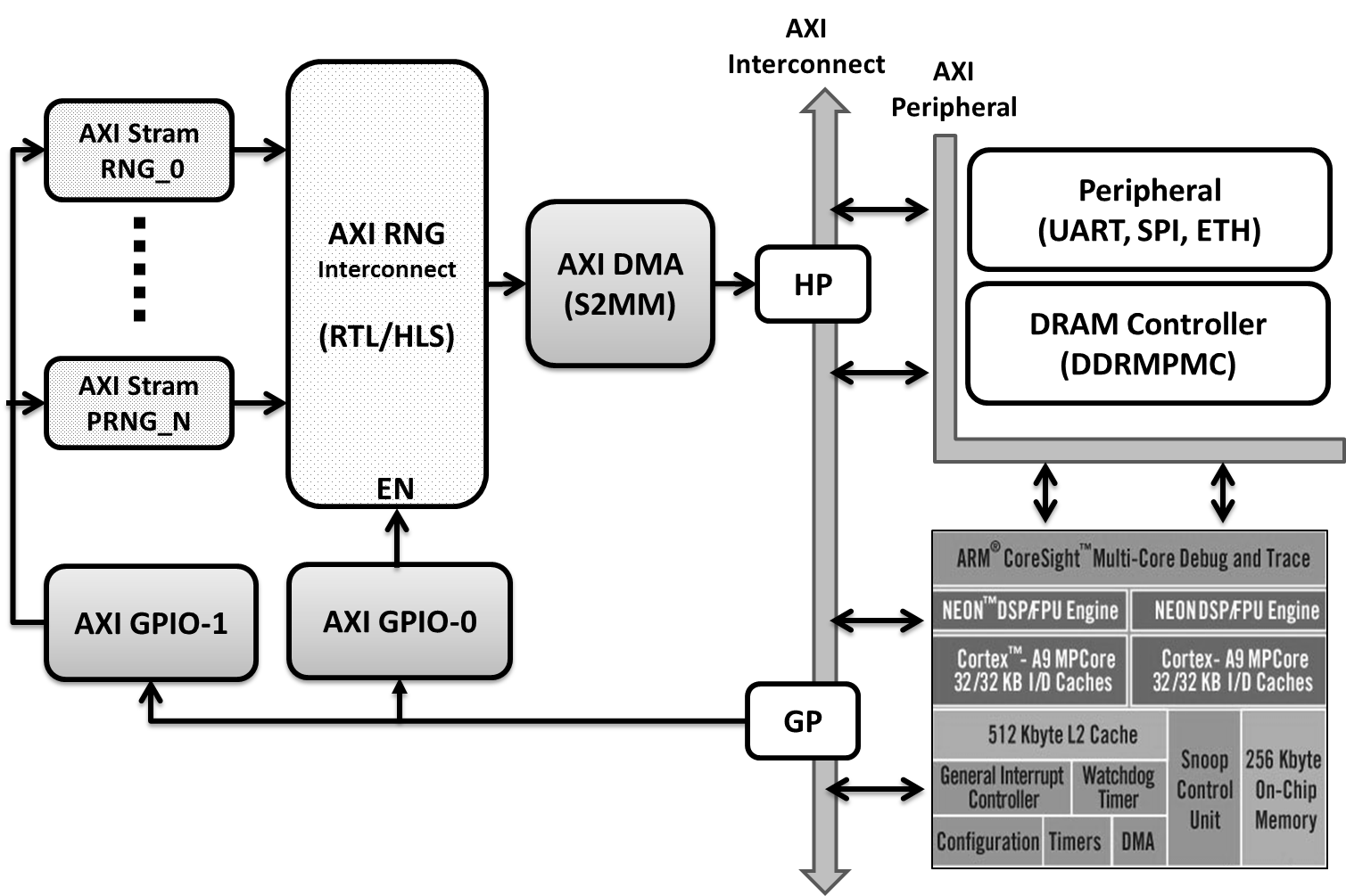}
  \caption{PRNG Platform Based on Zynq}
  \label{fig:zynq}
 \end{figure}

The hardware architecture of our system used to integrate and test PRNGs is illstrated in Figure~\ref{fig:zynq}. It contains, respectively: the ARM Cortex-$A9$ dual cores MPSoC, the high performance DDR3 $512$Mb, an UART, and finally the PRNGs (RTL or HLS implementation). Additionally, to read the  random output on the CPU, we have used both an AXI-PRNG interconnect and an AXI Direct Memory Access controller engine (DMA). The firmware for it parts, is used to initialize the system, for transaction synchronization, and for the interface with an external peripheral.

Meanwhile, the CPU initialises and reads/writes data of an IP in PL (\emph{i.e.}, PRNG) over the AXI master using  general-purpose GP ports. On the other hand, the AXI slave is used for PL master IP over High Performance (HP) ports. Each of these interfaces can handle up to $16$ bytes of data. The interface protocol, for its part, can be configured either as \emph{Stream} for high-speed streaming data, or as \emph{Lite/Full} for high-performance memory-mapped requirements (data transactions over an address). 
 
This interconnect component is re-configurable using the firmware, which deploys two GPIO IPs for that task. GPIO-$0$ is used to select one PRNG at a time, and GPIO-$1$ is used for the data burst size of the PRNG. 
For instance, all PRNGs implemented in HLS or RTL including the AXI-PRNG interconnect are \emph{AXI Stream Interface}, while the CPU is \emph{Memory-Mapped Interface}. 
Additionally to CPU, the AXI DMA engines, which oversees the data transaction between the slave and master IPs, deploys the receiver channel \emph{Slave to Memory Map} (S2MM) connected to a salve port and the transmitter channel \emph{Memory-Map to Slave} (MM$2$S) connected with the master. 

\subsection {Comparison} 
Table~\ref{Tab:hlsimp} and Table~\ref{Tab:rtlimp} give some performance results 
of PRNG implementation in terms of area (space) and throughput (speed). 
The Xilinx tool calculates all resources used in FPGA as logic gates, LUT, Flip-Flop (register), additionally to DSP and memory blocks. 
Hence, for our area comparison, we only calculated LUT and FF as $(LUT+FF)\times 8$, since DSPs and RAM memories are hard blocs that can mostly affect time performances. The throughput performance is calculated as Frequency$\times$ Output range. It depends on two parameters, namely the logic critical path used and the output range ($32$ or $64$ bits).

We obtained that the lowest area resources are for LUT-SR, Taus$88$, and xorshift$64$, while combined PRNGs like KISS and MRG$32k3a$ have a large area consumption too. 
Additionally, the throughput of Taus$88$ and LUT-SR with LFSR$113$ of $32$ bit generators, have the highest throughput performance, while the best are xorshift$64$ and LFSR$258$ in the $64$ bit case. On the other hand, the LCG and TGFSR families are expected to have the lowest throughput performance, as they operate large arithmetic operations like $64$ bit multiplications using DSP (it will be worse when using LUT). Besides that, using memories for TGFSR will drop the PRNG frequency automatically to the half without counting other logic. Once again, the combined generators have the weakest throughput performances. To conclude the FPGA resource performance aspects of this comparison, LFSR and xorshift PRNGs are more recommended to limit space and for better speed performances in hardware applications (mobile phone, smart cards, and so on).

Hardware PRNGs presented here must be evaluated too regarding their randomness, which can be done using statistical tests. 
The \emph{TestU01} battery is currently the most complete and stringent battery of tests for RNGs, which groups more than $516$ tests inside $7$ big sub-batteries. Among them, the \emph{Big Crush} is the most difficult one.

\begin{figure}[ht]
\centering
\includegraphics[width=0.45\textwidth]{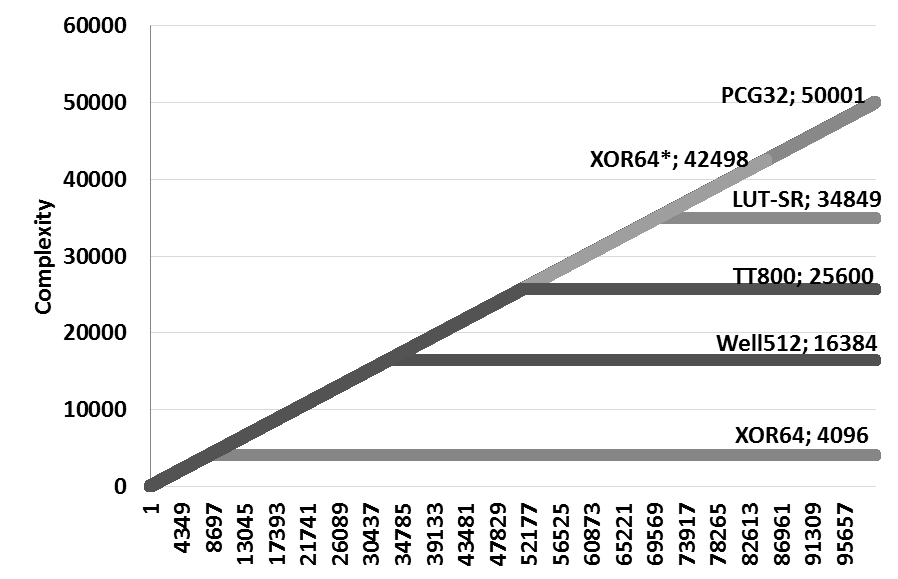}
\caption{Linear Complexity Test failing for TestU01}
\label{fig:xor64}
\end{figure}

After applying our experiments illustrated in Figure~\ref{fig:xor64}, we have obtained that only PCG$32$, MRG$32K3a$, and xorshift$^*$ generators can pass the Big-Crush of TestU$01$, which is coherent with the literature. Obtained test results have shown that a particular and common test called the linearity complexity test is very frequently failed. In details, TestU01 uses the \emph{Berlekamp-Massey algorithm} with the jump statistic to calculate the expected values compared to a chi-square test (the expected value).
Such a failure is related to what has been detailed in Section~\ref{cpmlex} about the linear complexity computation. Indeed all PRNGs are linear, but this does not lead to the linear complexity of a long random sequence.

To put it in a nutshell, if we take the ratio of area/throughput as main criterion, we are balancing between high performance (xorshift$64$ and LFSR$113$) and the ability to pass statistical tests (PCG$32$ and xorshift$^*$), which is not surprising. Another result is that combining PRNGs leads to a performance decrease in hardware level.
Next section studies a family of specific combinations which are based on \emph{Chaotic Iteration}.

%% file: 6-Chaos.tex
In this section, a recent pseudorandom number post treatment based on \emph{Chaotic Iterations} (CIs ~\cite{bgw09:ip,fang2014fpga,bahi2013fpga}) is recalled. It is based on Devaney~\cite{Devaney} theory of chaos.
This theory focuses on recurrent sequences of the form $x_0 \in \mathbb{R}$: $x_{i+1} = f(x_i)$, and studies for which function $f$ such sequences presents elements of complexity and disorder. In particular, it is wondered when effects of an alteration of the initial term $x_0$ can be predicted. Such chaotic sequences are candidate to provide pseudorandomness, leading to the field of chaotic pseudorandom number generators (CPRNGs). 

Let us now recall the mathematical definition of chaotic iterations CIs~\cite{bgw09:ip}. They are a particular kind of vectorial discrete dynamical system in which at $i$-th iteration, only a subset of components of the iteration vector are updated.

\theoremstyle{definition}
\begin{definition}
\label{defIC}
Let $f: \{0;1\}^\mathsf{N} \longrightarrow \{0;1\}^\mathsf{N}$ and 
$S \in \mathcal{P} \left(\llbracket1,\mathsf{N}\rrbracket\right)^\mathds{N}$
a sequence of subsets of the integer interval $\llbracket1,\mathsf{N}\rrbracket$ called a ``chaotic strategy'', where $\mathcal{P}(X)$ is the set of all subsets of $X$ and $\mathds{N}$ is the set of natural numbers. 
\emph{General chaotic iterations} $(f, (x^0, S))$ are defined
for any  $n \in \mathds{N}^*$ and $i \in \llbracket 1; \mathsf{N} \rrbracket$ by:
$$\left\{
\begin{array}{l}
x^0 \in \mathds{B}^\mathsf{N}, \ \mathsf{N}\geqslant 2 \\
  x^{n}_i = \left\{
\begin{array}{ll}
x^{n-1}_{i} & \textrm{if } i \notin S^n\\
f(x^{n-1})_{i} & \textrm{if } i \in S^n.
\end{array}
\right.
\end{array}
\right.$$
\end{definition}

For our PRNG applications, CIs have been implemented by the following process. The iteration function $f$ is the negation function ($f((x_1, \hdots, x_\mathsf{N}))=(\overline{x_1}, \hdots, \overline{x_\mathsf{N}})$). In this case, the CI based pseudorandom number generator is denoted by XOR-CIPRNG, which can be rewritten as $x_{i+1}=x_i \otimes S_i$~\cite{bcgh15:ij}. In the modified version we implemented, two inputted PRNGs denoted by $x_i$ and $y_i$ are used for defining the chaotic strategy $S$, as described in Algorithm~\ref{eq:ci}. Furthermore, we added a third inputted set generator $z_i$ for more complexity. This generator will pick randomly a subset of the inputs at each iteration. Only the $\log(\log(n))$ least significant bits (in this case, $3$ bits) are finally taken for pseudorandomness.

\begin{figure}[ht]
\begin{small}
\centering
\removelatexerror
  \begin{algorithm}[H]
  \DontPrintSemicolon
\caption{Xorshift based Chaotic Iteration}
\label{eq:ci}
\KwIn{$s$ (a 32-bit word)}
    \KwOut{$r$ (a 32-bit word)}  
    $X_i \gets PRNG1$,
    $y_i \gets PRNG2$,
    $z_i \gets PRNG3$\;
    \If{$(z_i\&1) \neq 0$}{$ s \gets s \otimes (x_i \& 0x0ffffffff)$\;}
    \If{$(z_i\&2) \neq 0$}{$ s \gets s \otimes (x_i \gg 32)$\;}
    \If{$(z_i\&4) \neq 0$}{$ s \gets s \otimes (y_i \& 0x0ffffffff)$\;}
    $r \gets s \otimes (y_i \gg 32)$\;
\end{algorithm}
\end{small}
\end{figure}

We tested more than $275$ combinations using CI post processing, a few of them being summarized in Table~\ref{tab:ci}.
In the first row of this table, triplets [$i, j, k$] represent the combination of PRNG1, PRNG2, and PRNG3 successively, where for $i$ and $j$, $0$ is for xorshift$64$, $1$ means xorshift$^+$, while the third component $k$ is respectively set to $1$,$2$,$3$,$4$, and $5$, corresponding to LFSR$113$, Taus$88$, TT$800$, WELLRNG$512$, and Mersenne Twister.

If we compare with the combined generators KISS and MRG$32k3a$ previously evaluated, we can notice the same characteristic in terms of area and throughput. Let us remark that some combinations need huge area resources, due to internal space required for some PRNGs like the Mersenne Twister or CMWC$4096$. But objective of this article is to show that PRNGs which previously failed some statistical tests can pass them after the CI post treatment: indeed, all the combinations of Table~\ref{tab:ci} achieve to pass the most stringent Big-Crush battery of Testu01. Furthermore, if we consider the combinations of [xorshift$64$, xorshift$^+$, LFSR$113$] or [xorshift$^+$, xorshift$^+$, Taus$88$], the obtained CI-PRNGs are more performing than MRG$32k3a$ (which also pass the TestU01) without using any DSP$\&$RAM blocs. 
To sum up, chaotic iterations post processing can contribute to increase the statistical performance of PRNGs.
\begin{table}[ht]
\centering
\caption{Chaotic Iterations Post Processing Implementation}
\label{tab:ci}
\begin{small}
\begin{tabular}{|L{1.2cm}|L{0.57cm}|L{0.57cm}|L{0.48cm}|L{0.57cm}|L{0.57cm}|L{0.48cm}|}
\hline 
PRNG    & 011   & 012   & 013   & 014   & 015   & 112   \\ \hline \hline
LUT     & 283   & 430   & 362   & 499   & 367   & 356   \\ \hline
FF      & 540   & 975   & 557   & 854   & 607   & 519   \\ \hline
DSP     & 0     & 6     & 0     & 3     & 2     & 0     \\ \hline
RAM     & 0     & 2     & 2     & 2     & 8     & 0     \\ \hline
Area/$10^3$  & 6.58   & 11.2  & 7.3   & 10.8  & 7.79   & 7.0   \\ \hline
T(Gbps) & 6.9   & 5.5   & 6.5   & 5     & 5.5   & 5.9  \\ \hline
\end{tabular}
\end{small}
\end{table}